\documentclass{aa}

\usepackage{graphicx}
\usepackage{txfonts}
\usepackage{natbib}
\usepackage{ulem}
\begin{document}

\title{Strength distribution of solar magnetic fields in photospheric quiet Sun regions}

\author{ {\sc J.C. Ram\'{\i}rez V\'elez} \inst{1}, {\sc A.~L{\'o}pez Ariste} \inst{2} and {\sc  M. Semel}\inst{1} }  
\institute{LESIA. Observatoire de Paris Meudon. 92195 Meudon, France; \email{Julio.Ramirez@obspm.fr, Meir.Semel@obspm.fr} \and
  THEMIS, CNRS UPS 853, c/v\'{\i}a L\'actea s/n. 38200. La
  Laguna, Tenerife, Spain; \email{arturo@themis.iac.es}}

\offprints{Julio.Ramirez@obspm.fr}

\begin{abstract}
{The magnetic topology of the solar photosphere in its quietest regions is hidden by the difficulties to disentangle magnetic flux through the resolution element from the field strength of  unresolved structures. The observation of spectral lines with strong coupling with hyperfine structure, like the observed Mn\,{\sc i} line at 553.7 {\rm nm}, allows such differentiation.}
{To analyse the distribution of field strengths in the network and intranetwork of the solar photosphere through inversion of the Mn\,{\sc i} line at 553.7 {\rm nm}.}
{An inversion code for the magnetic field using the
Principal Component Analysis \textit{(PCA)} has been 
developed. Statistical tests are run on the code to validate it.
The code has to draw information from the small-amplitude spectral feature appearing in the core of the Stokes V profile of the observed line for field strengths below a certain threshold, coinciding with lower limit of the Paschen-Back effect in the fine structure of the involved atomic levels.}
{The inversion of the observed profiles, using the circular polarization (V)
and the intensity (I), shows the presence of magnetic fields 
strengths in a range from 0 to 2 {\rm kG}, with predominant weak 
 strength values. Mixed regions with mean strength field values 
of 1130 and  435 Gauss 
are found associated with the network and intranetwork respectively.}
{The Mn\,{\sc i} line at 553 nm probes the field strength distribution in the quiet sun and shows the predominance of weak, hectoGauss fields in the intranetwork, and strong, kiloGauss fields in the network. It also shows that both network and intranetwork are to be understood at our present spatial resolutions as field distributions of which we hint the mean properties.}

\end{abstract}
\titlerunning{ Magnetic field strengths in quiet regions}
\authorrunning{Ram\' irez V\'elez  et al.\ \ }

\date{Accepted in A\&A}
\maketitle

%%%%%%%%%%%%%%%%%%%%%%
\section{Introduction}

The absence of observables of the magnetism of the quiet sun is the main handicap in the determination of the magnetic topology of this region covering most of the solar photosphere. Over-interpretation of the scarce information contained in the usually observed spectral lines in magnetometry, and its comparison with numerical simulations of magnetoconvection,  has been the subject of long disputes about the nature of those fields. The addition of further observables, able to constrain the models and to unveil hidden ambiguities and biases in the diagnostic techniques was mandatory in such a situation. This work is a contribution in that direction, with a further step forward in the use of Mn lines with a strong hyperfine coupling. 

In a short descriptive summary of the previous studies of the magnetism in quiet Sun regions,  we find from one side, the Fe lines  in the visible region whose circular polarization amplitudes are almost uniquely sensitive to magnetic flux, (e.g.  Keller et al. 1994; S\'{a}nchez Almeida \& Lites 2000; Lites 2002, Dom\'{i}nguez Cerde\~na et al. 2003; Orozco Su\'{a}rez et al. 2007),   and on the other side, we find  the inversions of the Fe lines in the near-IR domain   whose profiles are almost universally split by Zeeman effect (Lin 1995; Lin \& Rimmele 1999; Khomenko et al. 2003). Traditionally, studies based on the visible spectral lines concluded on the presence of fields with kG strengths and small surface coverage, while studies based on infrared lines favoured weaker hG fields and more spread in surface. 

Recently, the picture has been made more realistic by considering a continuum distribution function for magnetic field strengths at each resolution element in the quiet sun \cite{socas_SA_03}, instead of a single magnetic vector value. Additionally,  a  very interesting new approach has enriched the individual line analysis in quiet regions: the inversions of simultaneous and co-spatial observations of the Fe lines in the visible (630 nm) and in the near-IR (1.5 $\mu$m).  Two such studies have already been carried out  founding  divergent results. 
In the first of these works, Dom\'{i}nguez Cerde\~na et al. (2006)  have employed a model with three magnetic components ({\sc mismas}  model, S\'anchez Almeida \& Landi Degl'Innocenti 1996) to simultaneously invert the lines in both spectral ranges. The authors  retrieved field  distributions in the second and third magnetic components with peaks in the strong regime and they conclude that the kG fields contribution dominates the magnetic flux and energy transport.  In  the other of these studies, Mart\'{i}nez Gonz\'{a}lez et al. (2008) showed  that when  only the inversions of the Fe lines in the visible are considered, the magnetic distribution is dominated by strong strength fields (kG). However, when simultaneously near-IR and visible inversions were both considered, the kG contributions disappears  and the distributions  are mostly dominated by weak strength fields (hG). This last  conclusion is  in accord to the results obtained from near-IR inversions alone (see previous references). These results, suggesting a probable bias of the quiet sun Fe inversions when performed only in the visible at 630 nm had also been  found on numerical tests  \cite{bellot_coll_03, marian_fe_06}.  In the counterpart, a possible bias in the inversions of the IR alone, has been also argued suggesting that this lines are slightly sensitive to the strong strength fields, such that when  mixed  and not resolved magnetic structures of hG and kG are both present, the inferred strengths of hG are preferred over those of kG (e.g. S\'{a}nchez Almeida \& Lites 2000; Dom\' inguez Cerde\~na et al. 2006).  Such discussions and differing results illustrate the difficulty and ambiguity of the measurements attempted, and have prompted the search for more observables.

Apart from those Zeeman-based techniques discussed above, the description should be completed with  the Hanle effect diagnostics,  pioneered by  Stenflo (1982). A paradigm of  not structured and  turbulent fields, e.g.  \cite{manso04},  could result in  apparent contradiction with the tube-like structures used to describe Zeeman-based observations if one insists in the picture of a single vector magnetic field per resolution element.

Mn lines subject to strong hyperfine coupling appeared as an interesting observable to be added to the previous ones: their amplitudes are dependent on magnetic flux as the usual Fe lines in the visible regime, but under certain magnetic field regimes a new spectral feature appears in the Mn profile. Just the presence or absence of such a spectral feature allows the observer to determine the presence or absence of fields with strengths below a certain threshold thus adding a new observable on field strength to be compared to measurements with near-IR lines.

Quiet sun conditions limit the information on magnetic fields to circular polarization profiles, whose amplitudes stay too near to the usual signal-to-noise levels found in solar polarimetry. The amplitudes of the linear polarization  Stokes parameters use to be  below the noise level  at the typical spatial resolutions, and are thus useless for analysis purposes;  and intensity profiles are insensitive to the weak magnetic fluxes involved. Consequently one does not only need to add further spectral lines to the amount of observables, but also do it under observing conditions that guarantee very high signal-to-noise levels in polarimetry. We used the telescope THEMIS for that purpose in this work and we explored the capabilities of the Mn lines to provide further information on the quiet sun magnetic strength distribution. We simplified the formation of the Mn lines with a radiative transfer model using a Milne-Eddington atmosphere; the model  proves to be at the limit of present possibilities with yet too many of its parameters left undetermined by the available observables. We nevertheless prove through synthetic tests that we can ascertain the distribution of field strengths in strong and weak flux regions  (photospheric network and intranetwork respectively) with good accuracy, within the limits of the adopted Milne-Eddington modeling for the Sun.

With the results presented here, that show a field strength distribution with maximum number of occurrences at very weak fields (less than or around 100 G),  we propose in the conclusions a cartoon scenario that would be consistent with  both the Zeeman-based measurements (visible and near-IR) and the Hanle measurements.

In section 2, we review the signatures of the magnetic field strength in the line formation of the Mn\,{\sc i} profiles. In section 3, we describe  the inversion code used in this work to retrieve the strength and magnetic flux from the data, and  we test its capabilities with realistic noise conditions. The observations are described in section 4 and  the results presented in section 5. Conclusions and a final overview are included in the final section.

%%%%%%%%%%%%%%%%%%%%%%
\section{Hyperfine structure and the Stokes parameters}

The coupling of the nuclear 
angular momentum ($\vec{I}$) with the orbital total angular momentum
($\vec{J}_{LS}$) gives rise to  hyperfine structure ({\sc hfs}). 
The total angular momentum  $\vec{F} = \vec{J} + \vec{I}$ results in new non-degenerated atomic levels.
The apparent splitting of the atomic transitions due to this effect, and the consequent 
broadening of solar and stellar spectral lines has been discussed in several 
papers since the work of Abt (1952), and 
used to measure the abundances of different elements \cite{kur_93}. 

The most common observable for the analysis of (solar or stellar) spectral lines modified by hyperfine structure is  the broadening of the line profile. Such broadening however can be the result of others physical phenomena  as well: micro-turbulence,  Doppler broadening or just blended lines. Too often the ambiguity on the origin of an observed line broadening cannot be solved for, even in observations with very high spectral resolution, unless full  spectropolarimetry is performed. Such is the case presented in this work, the Mn\,{\sc i} line at 553.7 nm: although producing a slightly asymmetric intensity profile, the recognition of the {\sc hfs} features becomes straightforward in circular polarisation, since the {\sc hfs}  together with  an atmosphere permeated of an  external magnetic field produce a unique spectral signature  in the Stokes V profiles.  The  use of this signature and its variations under different conditions as a tool for the detection  of magnetic fields in the quiet sun in different field strength regimes has been theoretically first proposed in L\'opez Ariste et. al (2002). Later, in L\'opez Ariste et. al (2006), the authors  showed that observationally it is possible to detect the {\sc hfs} effect in the circular polarization profiles.
In this work we extend previous studies and we build an inversion code drawing its information from those spectral features in the Stokes profiles due to {\sc hfs}. 
The inversion algorithm is firstly validated and characterized with synthetic data, and then applied to observed data. A quantitative analysis of the field strength distribution in the quiet sun results.

\begin{figure}[ht]
\resizebox{9.cm}{!}{\includegraphics{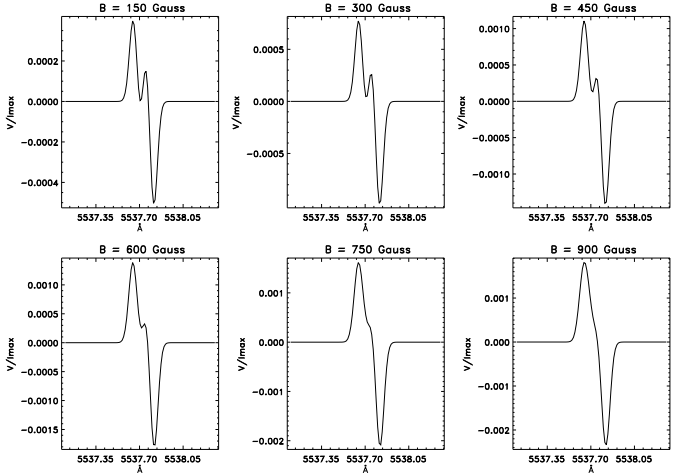}}
\caption{Stokes V profiles  as a function of the magnetic field strength. 
Note that the spectral feature on the central part of the profiles
is the signature associated to 
the {\sc hfs} regime, and that it  
disappears about B $\sim$ 750 Gauss, as the line passes into the incomplete
Paschen-Back regime.}
\label{fig:chipote}
\end{figure}

For illustration and reference we show, in Fig. \ref{fig:chipote}, the calculated  spectral signature of the Stokes V profiles for the Mn\,{\sc i} at 553.7 nm, as a function of the magnetic field strength. The important remark from the V profiles  pictured is that the weaker the magnetic field strength, the higher the amplitude of the spectral feature due to the {\sc hfs}, visible in the central part of the profile.

%%%%%%%%%%%%%%%%%%%%%%
\section{The inversion code}
%\subsection{Generation of the database}
\subsection{Solution to the radiative transfer equations}
The  four coupled equations for the radiative transfer of the Stokes 
parameters %\ref{equ:RT}
were solved analytically  by the code {\sc diagonal} 
(L{\'o}pez Ariste \& Semel 1999),
which was slightly modified from its original version
to include the computation of hyperfine structure.

To use the code {\sc diagonal} we assume a Milne-Eddington atmosphere, with the possibility to include numerically gradients with optical depth of the velocity and magnetic fields, and adding up to fifteen free parameters. Evidently, the use of a bigger number of free parameters implies a higher risk of inversion ambiguities and biases, if they happen to exceed the dimensionality of the observables.  But on the other hand considering less parameters may not allow to cover all the variety of the observed profiles. We decided thus to limit the number of parameters for each atmospheric model, to the ones described below, leaving out the velocity and magnetic gradients along the line of sigth.%, responsible for  asymmetries in the profiles.}

The inversion code works in hierarchical mode by a double pass fit. In the first pass only the intensity is fitted, permitting half of the atmospheric parameters (Doppler  width, $V_D$, velocity in the line of sight $V_{LOS}$, $\eta_0$ and the source function gradient, $\nabla\tau $) to be constrained in their ranges of variation. The remaining four atmospheric parameters are related to the presence  of a magnetic field: the angle between the magnetic field direction and the direction of the line of sigth $\theta_{BLOS}$, the azimuthal angle $\theta_z$, the strength of the magnetic field $B$ and a filling factor $ff$, accounting for unresolved structure. In the second (and final) fit, the eight parameters are considered in fitting simultaneously the profiles of intensity (I) and circular polarization (V). Each atmospheric parameter will be left to vary randomly in the range of values specified in table \ref{tab:tab1} in order to build a database for the inversions.

%%%%%%%%%%%%%%%%%%%%%%%%%%%%%%%%%%%%%%%
\subsection{Atmospheric models and database}

Let $\vec{\chi}$ denote  a given combination of the described atmospheric
parameters such that,

\begin{equation}
\vec{\chi}_m= \left( V_D,V_{LOS},\eta_0,\nabla\tau,B,
\theta_{BLOS},\theta_z,ff\right)_m
\label{equ:chi}
\end{equation}
is the $m_{th}$ combination.
Let $\vec{S}_m(\vec{\chi}_m)$ denote
the Stokes vector calculated for this combination of
atmospheric parameters $\vec{\chi}_m$:

\begin{equation}
\vec{S}_m(\vec{\chi}_m) = \left( I_m(\vec{\chi}_m),
Q_m(\vec{\chi}_m),U_m(\vec{\chi}_m),
V_m(\vec{\chi}_m) \right),
\label{equ:P}
\end{equation}
and let  $P_m$ denote any of the components of $\vec{S}_m(\vec{\chi}_m)$.

In the  computation Stokes parameters we have considered  201 spectral wavelength points with a step size of 12 $m \AA$,  covering  in total 2.41 $\AA$ around the line center  at 553.762 nm. In the following, we proceed to consider each profile ${P}_m$ as a vector of 201 points $\vec{P}_m$.
The creation of the database is the most important step for the inversions  working with principal components analysis (e.g. Rees et al. 2000; Socas-Navarro et al. 2001).  It is required to consider a large enough$^1$ number of profiles  in order to consider the database as statistically representative. Thus, firstly, we have created a database of 80802\footnote{The actual figure has no particular meaning other than illustrating what we consider as \textit{large enough} and for future reference.}  profiles $\vec{P}_m$, hereafter called the profile space matrix  ($\mathcal{M}$), and  we assume that all possible observed profiles  in the data set are represented in the database. This assumption is however verified in the tests presented at the end of this section.

We next perform a \textit{Singular Value Decomposition (SVD)} \cite{golub_96}  of the space matrix to compute an algebraic basis of the space spanned by the profiles. We should refer to the elements of the basis as eigenprofiles. 

%%%%%%%%%%%%%%%%%%%%%%
\begin{table*}[htpb]
\caption{All the parameters vary randomly in the indicated ranges.
The first four  parameters of the \textit{initial set},
are used to fit the intensity in the first inversion passage in the code.
In the \textit{final set}, all the
eigth parameters are considered to invert simultaneously (I) and (V).}

\label{tab:tab1}
\centering
\begin{tabular}{ccccccccc}
\hline
Parameters & $V_D$ (m\AA) & $\eta_{0}$ & $\nabla \tau$  &  $V_{LOS}$ (km/s)  
& $B$ (G) & $\theta_{BLOS}$ (deg.) & $\theta_{z}$ (deg.) & ff \\
%\hline
%  & (km/s) & & & (km/s) & (G)& (deg.) &(deg.) \\
\hline
%\hline
\textit{Initial Set}   & [35,75] & [0,2.5] & [0,15] &  [-4,4]  &
[0,2000] & [0,90]  & [0,90]  & [0,1] \\
%\hline
Inverted (max,min) & (41,60) & (0.7,1.8) & (1.0,14.9) &
(0,0.54)  & --- & --- & --- &  --- \\
%\hline
mean value  &  47  & 1.16  & 8.8 & 0.02 & --- & --- & ---  &  --- \\
%\hline
\hline
\textit{Final Set}  & [39,61] & [0.5,2] & [0,15] &
[-2,2]&  [0,2000] & [0,90] & [0,90]&  [0,1] \\
\hline
\end{tabular}
\end{table*}
%%%%%%%%%%%%%%%%%%%%%%%%%%%%%%%%%
\subsection{Analysis with PCA}

Following the synthetic notation of Eq. (\ref{equ:P}),
for each one of the Stokes parameters ($I,Q,U,V$) it is possible 
to construct the $201 \times  80802$  profile matrix $\mathcal{M}$
that contains all the profiles $\vec{P}_m$ of the database:
\begin{equation}
\mathcal{M}= \left( \vec{P}_0  \ \ \cdots \ \
\vec{P}_{80801} \right)^T.
\label{equ:M}
\end{equation}
 After computation of the eigenprofiles, each profile in the
matrix  $\mathcal{M}$ can be reconstructed as:
\begin{equation}
\vec{P}_{m} (\vec{\chi}_m) =\Sigma_i \  \alpha_i(\vec{\chi}_m)  \  \
\hat{\vec{e}_i}  \ \ \ \ ;  i=0,1,...,200
\label{equ:p_pca}
\end{equation}
where $\hat{\vec{e}}_i$ are the eigenvectors associated to
the space profile ($\mathcal{M}$) and
$\alpha_i$ are the coefficients of the linear combination 
associated to the profile $\vec{P_m}$. Since the
eigenvectors form an orthonormal basis of the space
($\mathcal{M}$) arranged in order of importance, one  can
cut the expansion  in Eq. (\ref{equ:p_pca}) and
preserve only the first eigenvectors that contain relevant
information on the shape of the profile. The rest of eigenvectors
are associated with the noise and there is no interest in preserving 
them \cite{hansen_92, rees_00}.
In our study we find that the appropriate number of components in  
Eq. (\ref{equ:p_pca}), is $i_{max}=30$.

With the help of the set of eigenvectors, $\{\hat{\vec{e_i}} \}$,
we create a second database ($\mathcal{M}^{\prime}$) that is independent 
from the ($\mathcal{M}$) matrix, used to compute the 
$\{\hat{\vec{e_i}} \}$. The ($\mathcal{M}^{\prime}$) database 
will be decomposed in terms of the coefficients $\alpha_i$ as follows.
Let us denote $\vec{P}_{n}$ a  vector
of the new (second) database. Using the  Eq. (\ref{equ:p_pca}),
\begin{equation}
\alpha_i(\vec{\chi}_n) = \vec{P}_n  (\vec{\chi}_n)\cdot \hat{\vec{e}_i}
\ \ \ \ ;  i=0,1,...,29.
\label{equ:alfas}
\end{equation}
In this way, each profile $\vec{P}_n$ (observed or calculated)
has associated  a  unique set of coefficients
$(\alpha_0,\ ... \ ,\alpha_{imax})$,  denoted hereafter
$\vec{\alpha}_n \equiv (\alpha_0,\ ... \ ,\alpha_{imax})$.
So instead of working in the profile matrix  space  ($\mathcal{M}^{\prime}$) 
it is equivalent, and faster,  to work in the  coefficient's 
space ($\mathcal{C}$).
From  Eq. (\ref{equ:M}),
and in analogy with  ($\mathcal{M}^{\prime}$),
we obtain for each profile $\vec{P}_n$ the correspondent
coefficients $\vec{\alpha}_n$
in order to build the  30 x 80802 coefficient matrix
($\mathcal{C}$),

\begin{equation}
\mathcal{C}= \left( \vec{\alpha}_0 \ \   \cdots  \ \
\vec{\alpha}_{80801} \right)^T.
\label{equ:C}
\end{equation}

%%%%%%%%%%%%%%%%%%%%%%%%%%%%%%%%%%%
%%%%%%%%%%%%%%%%%%%%%%%%%%%%%%%%%%%
\subsection{Description and  validation of the inversion code}
One  important quality of the code developed  is that it is very easy to  estimate the error bars per model parameter. We present here the tests applied for that purpose before pursuing into the inversion of the observed profiles.

Let   $\vec{P}_{syn}$ denote a
synthetic profile with associated  coefficients   $\vec{\alpha}_{syn}$.
The profile  $\vec{P}_{inv}$, considered  to be the solution by the 
inversion code, is found in the space of coefficients ($\mathcal{C})$ as the 
nearest  set of coefficients (in an Euclidean sense) to  $\vec{\alpha}_{syn}$ :
\begin{equation}
 \| \vec{\alpha}_{inv} - \vec{\alpha}_{syn}\| = min \left( \| \vec{\alpha}_{syn} - \vec{\alpha}_{n}\| 
\right) \ \ \ ; \ \  n=0,1,...,80801.
\label{equ:coefs}
\end{equation}

Since  each one of the Stokes parameters would give, if considered separately, one solution to the Eq. (\ref{equ:coefs}), let us say ($\vec{\alpha}_{inv,I},\vec{\alpha}_{inv,Q}, \vec{\alpha}_{inv,U},\vec{\alpha}_{inv,V}$), it is  required to find the best solution for the four Stokes parameters simultaneously. For this purpose and given that it is a well-known fact, but particularly true in our data sets, that $I_n \gg  V_n > U_n,Q_n$  necessarily implying that, in average, $\vec{\alpha}_{inv,I} > \vec{\alpha}_{inv,V} >  \vec{\alpha}_{inv,U},\vec{\alpha}_{inv,Q}$  (e.g. L\'opez Ariste \& Casini 2002),  we  find  useful to introduce a \textit{weight}  for each solution. Thanks to this, the polarized signals are wighted similarly  to the intensity one, despite their disparate amplitudes.

Let the scalars
\begin{equation}
{\alpha}_{w,S} =  \frac {max( \vec{\alpha}_{syn,I})} 
{max(\vec{\alpha}_{syn,S})}
 \ \ \ \ \  \ ;  S=I,Q,U,V
\label{equ:coefsol}
\end{equation}
be the  factors that modify the computation of the norms  in Eq. (\ref{equ:coefs}), such that each one of the four solutions has the appropriate \textit{weight} and can be added to the others in a linear way. The final solution, $\vec{\alpha}_{sol}$, is then found as: 
\begin{equation}
\left( \Sigma_S {\alpha}_{w,S} \| \vec{\alpha}_{inv,S} - \vec{\alpha}_{syn,S} \|\right)  = min   \left( \Sigma_S {\alpha}_{w,S} \| \vec{\alpha}_{syn,S} - \vec{\alpha}_{n,S}\| \right) 
% \ \ \ \ ;  S=I,Q,U,V.
\label{equ:sol}
\end{equation}
where $S$ denotes, as usual, the Stokes parameters (I,Q,U,V).

Since we know from observations, described in section 4, 
that the signal-to-noise ratio  in  linear polarization
(Stokes $Q,U$) is not big enough to include them in the analysis,
they were excluded as well from the tests applied to the code  and
only the intensity and circular polarization  were considered. 
This implies that from Eq. (\ref{equ:coefsol}), $\  \ S=I,V$.

Finally, we  compare the solution values of  $\vec{\chi}_{sol}$  (associated to $\vec{\alpha}_{sol}$) to the known values of $\vec{\chi}_{syn}$ (associated to $\vec{\alpha}_{syn}$).  This comparison leads  to the statistical determination of the inversion errors per model parameter, where by error we mean the absolute difference between the original and the retrieved values.

%%%%%%%%%%%%%%%%%%%%%%%%%%%%%%%%

%%%%%%%%%%%%%%%%%%%%%%%%%%%%%%%%%%%%%%%%%%%%%%%%%%%%%
\subsubsection{Discussion of the tests}

In order to not bias the tests we have constrained the thermodynamic parameters to the ranges of variation retrieved from the observational data sets after the first inversion (see the first three rows of table \ref{tab:tab1}). Otherwise, we have not imposed any \emph{a priori} constrain to the magnetic parameters, and the profiles used in the tests were synthesized using independent uniform distributions  for each parameter, following the values found in the last row of table \ref{tab:tab1}. However, in the search of direct translation to the inversion of quiet sun profiles, we have conditioned the profiles to have amplitudes in the V profiles that do not exceed the maximum amplitude found in the data set ($ V < 5 \times 10^{-3}$) while being bigger than the lower noise level ($V> 5 \times 10^{-5} $). Under these conditions, it is expected that the sample of   profiles used in the tests is comparable with that from quiet sun data. 

\begin{figure}[ht]
\resizebox{9.5cm}{!}{\includegraphics{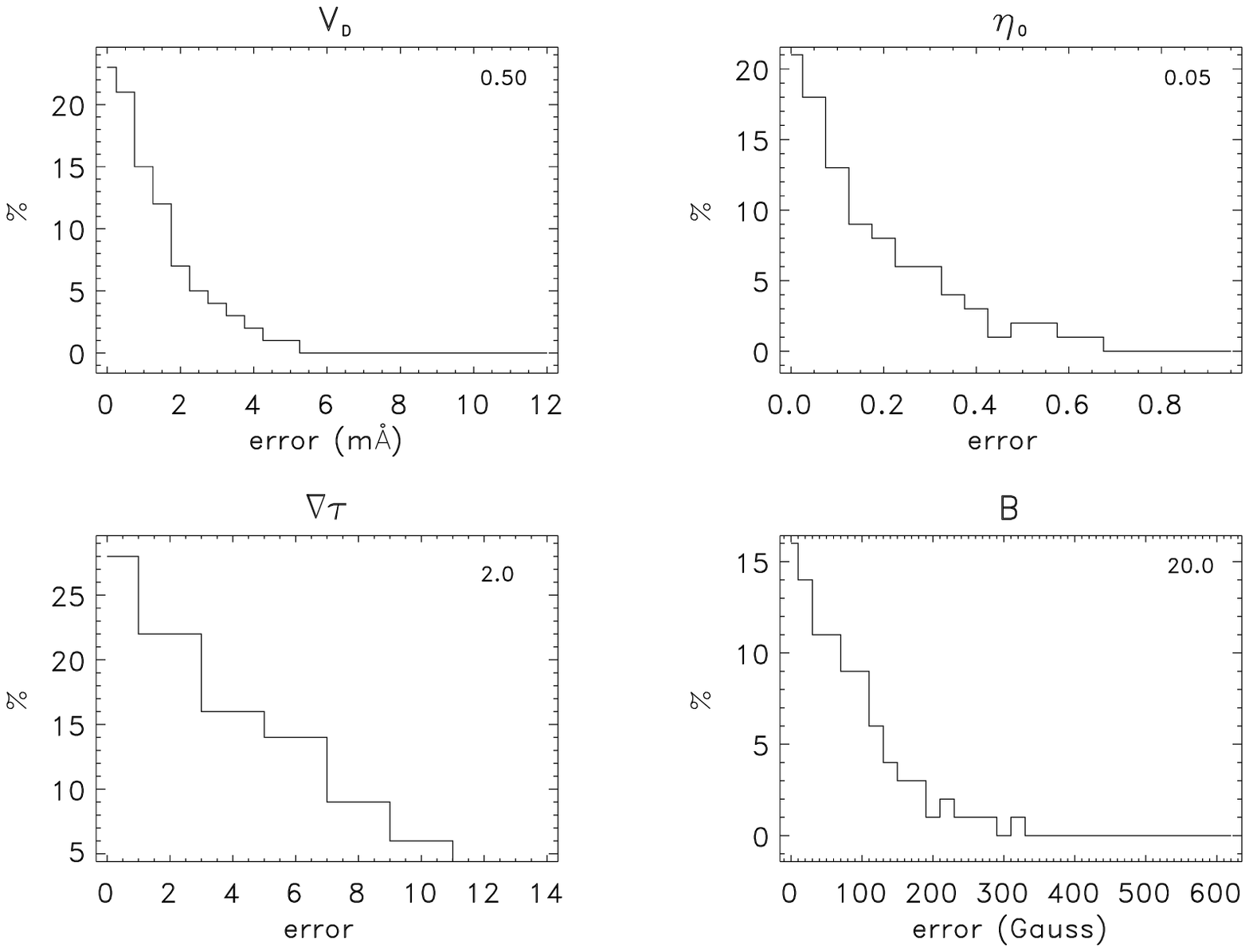}}
\resizebox{9.5cm}{!}{\includegraphics{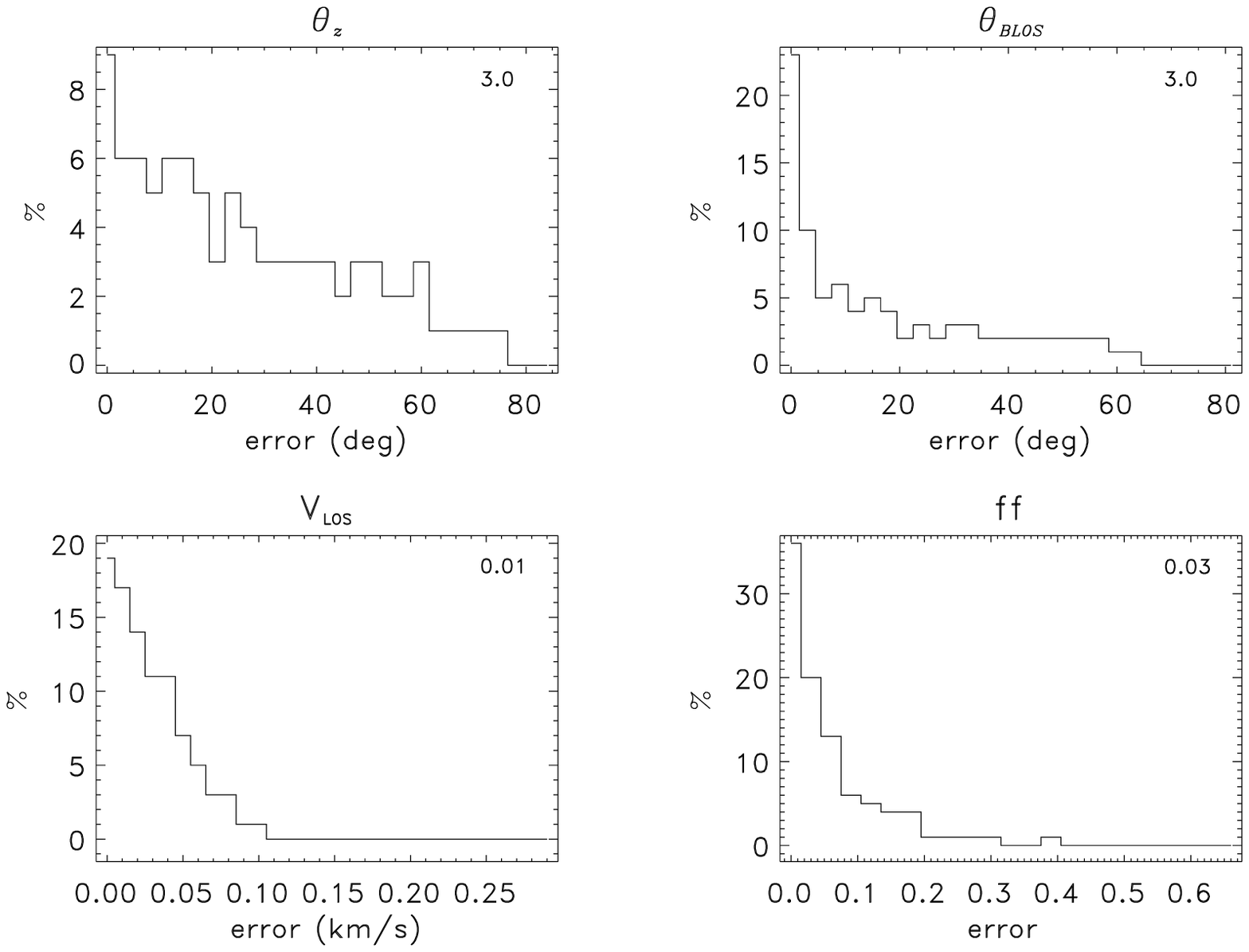}}
\caption{Inversions histograms of each atmospheric parameter. The number
in the upper right corner indicates the value of the bin size used.}
\label{fig:histos}
\end{figure}

In Fig. \ref{fig:histos}, we show the results of the inversion 
of samples of 500 profiles. As a first test, we have not included noise 
in the inverted profiles. With the exceptions of  $\nabla \tau$ and $\theta_z$, the  errors in  most of the parameters decrease rapidly to zero. 

 The bad results for $\theta_z$ are due to the fact that the linear polarization, Stokes $Q,U$, was not included in the inversions,   implying that the inverted values in this parameter correspond to  an aleatory distribution.  Consequently, the original (by construction) and the inverted distributions  in the azimuthal parameter are aleatory.  Finally, plotting the absolute difference between these two distributions results in an histogram with a pick error centered in (around) zero and that decrease smoothly, like the one obtained for $\theta_z$.

Now, the most relevant parameter in our analysis is the magnetic field
strength $B$, since the major interest of this work is to
analyse the magnetic field  in quiet sun conditions.  In what follows
we focus the discussion on this parameter and we include 
also the results of inversion of profiles with  noise. 

\subsubsection{Analysis of effective inversions in B}
\begin{figure}[ht]
\resizebox{9cm}{!}{\includegraphics{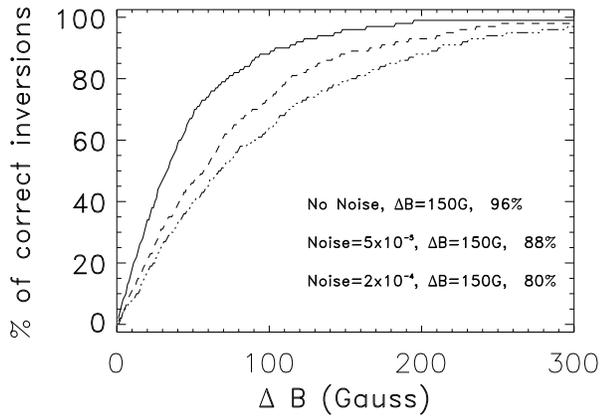}}
\caption{Curves of cumulative percentage of correct inversions. 
Each sample consisted of 500 profiles.
The solid line corresponds to  profiles without noise. The dashed line 
corresponds to a noise level of  $5x 10^{-5}$ and the dash-point-dash line 
to a noise level of $2 \times 10^{-4}$ the intensity of the continuum. 
The inset indicates the efficiency percentage 
for an upper limit of $\Delta B= 150$ G.}
\label{fig:testb}
\end{figure}

The graphics  in Fig. \ref{fig:histos} correspond  
to the ideal case of profiles without noise. Considering 
that the real profiles have a typical noise level of the order of
$2 \times 10^{-4}$ to $5 x 10^{-5}$ the continuum level, we proceed to
invert synthetic profiles with noise added to those levels. 
Let $B_{inv}$ denote the retrieved value of the field strength, and let 
 $B_{syn}$ denote the real atmospheric value of the synthetic profile. 
Representing by ($\Delta B$) the upper limit of the error value, 
such that 

\begin{equation}
\Delta B = | B_{inv} - B_{syn} |,
\end{equation}
we can characterize the inversion efficiency behaviour of the code,
understood as the percentage of profiles correctly inverted as a function
of ($\Delta B$).
In table \ref{tab:tab2}, we compare the 
inversions of the magnetic strength for profiles with different noise levels, 
and in Fig. \ref{fig:testb} we plot the respective cumulative 
distribution functions.

\begin{table}[ht]
\caption{Inversions efficiency  of the magnetic field strength,  expressed in percentage of correct inversions as function of $\Delta B$ .}
\label{tab:tab2}
\centering
\begin{tabular}{c c c c c c c}
%\hline
\hline\hline
$\Delta B$ &&&  \multicolumn{3}{c}{Noise level}    \\
 &&&  0 & $5 x 10^{-5}$ & $2 x 10^{-4}$ \\
\hline \hline
$ 50$ G &&& 69 \% & 46 \% & 40  \% \\
$ 100$ G &&& 89 \% & 74 \% & 65 \%   \\
$ 150$ G &&& 96 \% & 88 \% & 80 \% \\
$ 200$ G &&& 99 \% & 93 \% & 89\%  \\
$ 300$ G &&& 100\% & 98 \% & 96 \% \\
\hline
\end{tabular}
\end{table}

The value of the magnetic field strength goes from 0  through 2000 G 
covering all the range of magnetic strengths inherent to the
quiet Sun regime. Since the spectral
signature in Stokes V profiles evidence of {\sc hfs} regime,
like those in Fig. \ref{fig:chipote}, disappears for values around
$B \sim  750$ G, we have special interest in
the behaviour of the code while
inverting profiles in the weak field regime.
In table \ref{tab:tab3} we show that the precision inverting 
the magnetic strength is mainly independent of the magnitude
of the magnetic field. The  inversion gently improves
with higher values of the field strength, but 
in general the precision in the result is preserved.
\begin{table}[ht]
\caption{Inversion's efficiency percentages in the weak field 
regime as function of the magnetic field strength.} 
\label{tab:tab3}
\centering
\begin{tabular}{ccccccccc}
%\hline
%\hline
$\Delta B$ &  B $\le$150G &B $\le$350G &B $\le$550G &B $\le$750G &  Noise  \\
\hline
\hline
% /themis/hfs/pdif/wmcc_vmax_vmin/wmmc_poid2.0.sav
$ 100 G$ & 75 \% & 77 \% & 81 \% & 84 \% &  No\\
$ 200 G$ & 98 \% & 99 \% & 99 \% & 99 \% &\\
\hline
$ 100 G$ & 60 \% & 57 \% & 61 \% & 64 \% & $5 x 10^{-5}$ \\
$ 200 G$ & 85 \% & 87 \% & 89 \% & 90 \% &\\
\hline
$ 100 G$ & 56 \% & 53 \% & 54 \% & 57 \% & $2 x 10^{-4}$\\
$ 200 G$ & 76 \% & 80 \% & 82 \% & 83 \% &\\
\hline

\end{tabular}
\end{table}
We conclude that, assuming a Milne-Eddington model, the  code developed in this work for the line of Mn\,{\sc i} at 553.7 nm, correctly inverts  profiles in very different field strength regimes: from very weak magnetic fields (lower than 100 G) up to 2 kG, with an error bar of 200G with 89 \% of probability (300G  with 96 \% of probability) independent of the field strength.
With such results we can assert that, even if the actual field strength cannot be measured from inversions with sufficient precision, at least the inversion algorithm can disentangle  without any ambiguity  the hG  from the kG regimes of field strength.

%%%%%%%%%%%%%%%%%%%%%%
\section{Observations}
We carried out two campaigns of observations at the THEMIS telescope \cite{LA_R_S00, gelly_06} in the multiline observation mode MTR. During the first campaign in June 2005, we searched for  the spectral signature of the hyperfine structure of the Mn\,{\sc i} atom among four candidate lines in the visible spectral region at wavelengths 539.4 nm, 547.0 nm, 551.6 nm and 553.7 nm, respectively. From the data, we concluded that out of these four lines, only the one at 553.7 nm  presented a detectable spectral signature of {\sc hfs} in the Stokes $V$ profiles clear enough to be used for diagnostics. During the second campaign in June 2006 we focused on this line. In both campaigns the Fe\,{\sc i} doublet at 630.1 and 630.2 nm  was observed simultaneously with the  Mn\,{\sc i} lines. Some relevant  results concerning the importance of simultaneously observing spectral lines from both atomic species, as well as a brief description of the data acquisition and reduction,  has been presented in \cite{hfs_LMR_06}.

In both  campaigns we scanned the solar disk center and, in at least one day per campaign, we enjoyed excellent seeing conditions in the early morning. We have used the polarized signals levels of the Fe doublet (630.2 nm), observed simultaneously and co-spatially,  to  verify  that  around the  scanned  area there were no traces of active regions, nor enhanced magnetic areas: the Stokes V amplitudes of this magnetically sensitive line stay below 1\% polarization at all times, indicative of mostly quiet photosphere, both network and intranetwork regions alike. The data sets consisted of temporal sequences of 100 modulation cycles in polarization. Each modulation cycle was made of 6 images with exposure times of either 300 ms or 500 ms.  The slit was kept fixed over the solar disk, up to residual seeing-induced image movements. In 2006 the new THEMIS correlation tracker was fully functional, thus further diminishing any residual movement. The total time of  these sequences, around 8 minutes, is comparable to the lifetime of the granulation. The dimensions of the slit spanned a region of 48 arc sec in two separate regions of 15.5 arc secs each, with a masked region of 17 arc sec between them (this masked region is a requirement of the polarimetry mask used for the observation, which places the splitted, orthogonal polarisation, beams 16 arcsec away one from the other in the direction along the slit). The slit width was kept at 0.5 arc sec resulting in a spectral resolution of 20 m\AA{} in agreement with the pixel spectral sampling of 10 m\AA{}.

We reduced the data with the {\em DeepStokes} software tool currently integrated to the THEMIS system facilities, modified to preserve spatial resolution. As delivered the \textit{DeepStokes} code sums all images in the temporal series to increase the signal-to-noise ratio. Because of the long-time integration, image motion and blurring results in a degradation of the image quality. Trying to avoid such a degradation of the spatial resolution, otherwise attained in the individual images of the time series, we modified the code so that the addition of further images in the time series was not done regardless of a further degradation in the image quality. 

Specifically, in order to decide if the modulation cycle $i+1$ was to be added to the sum of the previous $i$ modulation cycles, a cut along the slit (at a pre-specified continuum wavelength of the intensity image)  for the  modulation cycle $i+1$  is compared to an analogous cut in the previous added   $i$ cycles. The number $N$ of points which differed in both such cuts in less than $D$ was measured. When $N$ happened to be bigger than a pre-specified threshold $N_0$ the $i+1$ cycle was deemed \textit{spatially coherent} with the previous ones and its Stokes parameters added to the series, increasing the signal-to-noise ratio. If the $i+1$ cycle failed to pass the test, it was assumed that the slit had moved and a new time series was started at point $i+1$. The sensitivity threshold $D$ and the minimum number of cases $N_0$ are provided by the user. We determined empirically their value from a posteriori considerations on attained signal-to-noise ratios and spatial resolutions, and the trade-off among both parameters. In the cases under analysis in this work  the achieved noise levels laid  between $5x10^{-5}$ and $2x10^{-4}$ of the continuum level, for a spatial resolution  better than 1 arc sec and approaching at times 0.6 arc secs\footnote{The spatial resolution was gauged by the crude and simple method of counting and measuring the size of brighter-than-average and darker-than-average regions as visible in the continuum cuts along the slit.}.

\section{Results}

\begin{figure*}[htpb]
\begin{center}
\resizebox{16cm}{!}{\includegraphics{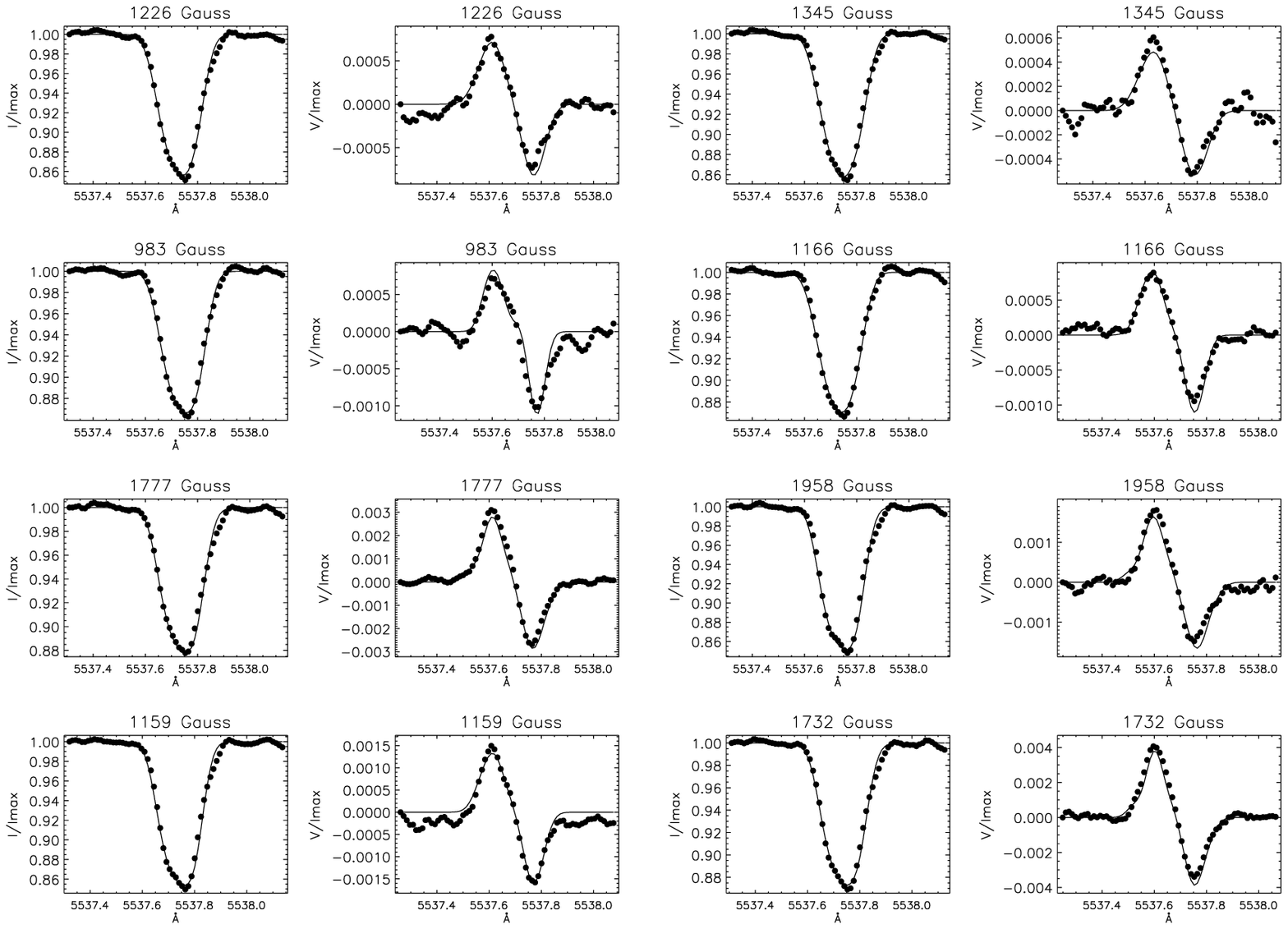}}
\resizebox{16cm}{!}{\includegraphics{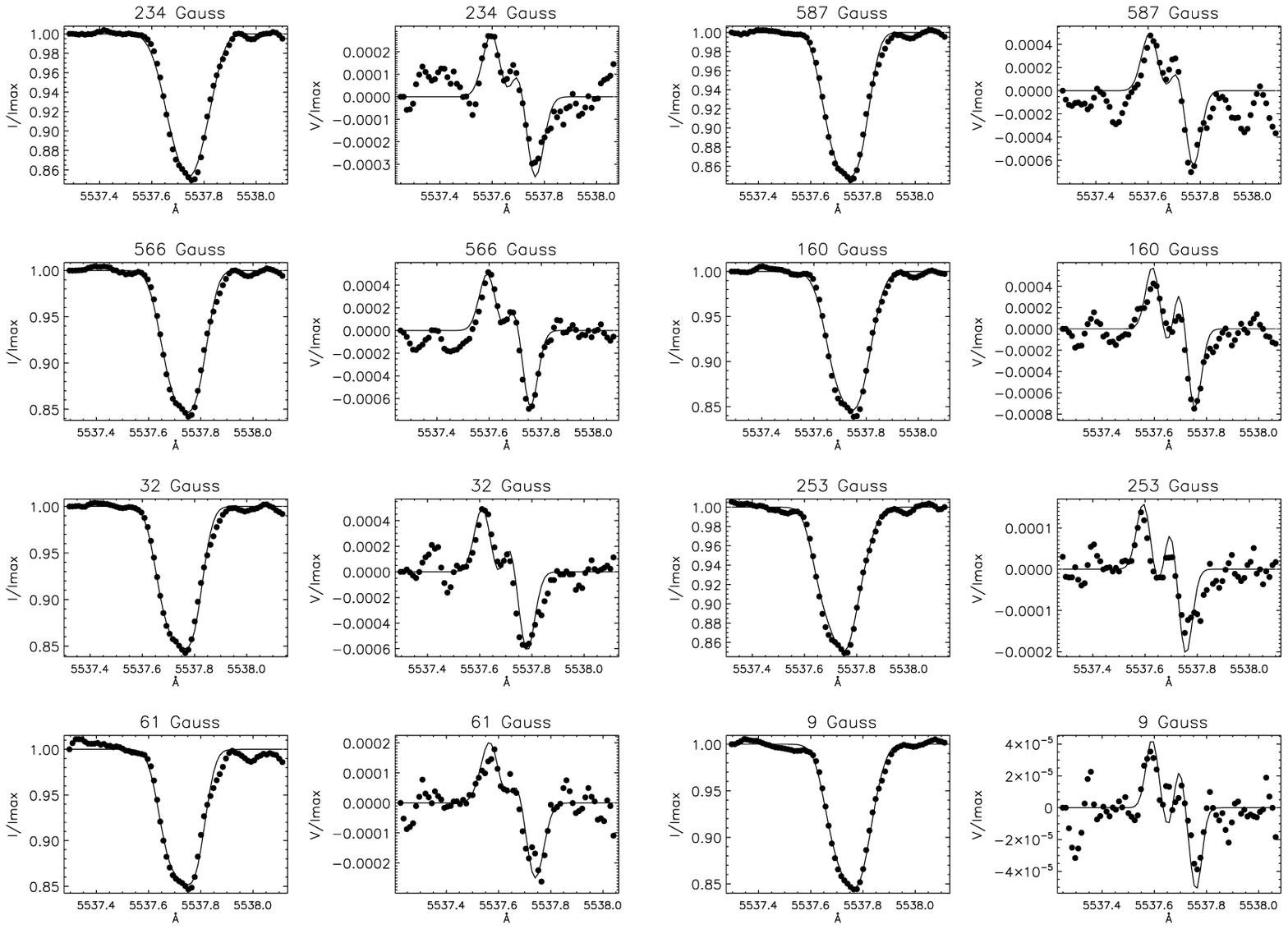}}
\caption{Examples of  inverted profiles. The dotted 
lines correspond to the observed data and the solid line to the respective
fitted profile. The profiles were inverted without any smoothing, but 
in this plot a 2-pixel smoothing has been applied to observed data.}
\label{fig:vperf}
\end{center}
\end{figure*}

In Fig. \ref{fig:vperf} we show for illustration some of the observed Stokes I and V
profiles  (dotted lines) and the correspondent fits from
the inversion (solid lines) overplotted.  The panel covers the diversity of  magnetic field strengths
found in the quiet sun as seen through the Mn\,{\sc i} lines: from the weak intensity regime (hG) to the strong regime (kG).

\subsection{The complete set of profiles}
\begin{figure*}[!htb]
\resizebox{9cm}{!}{\includegraphics{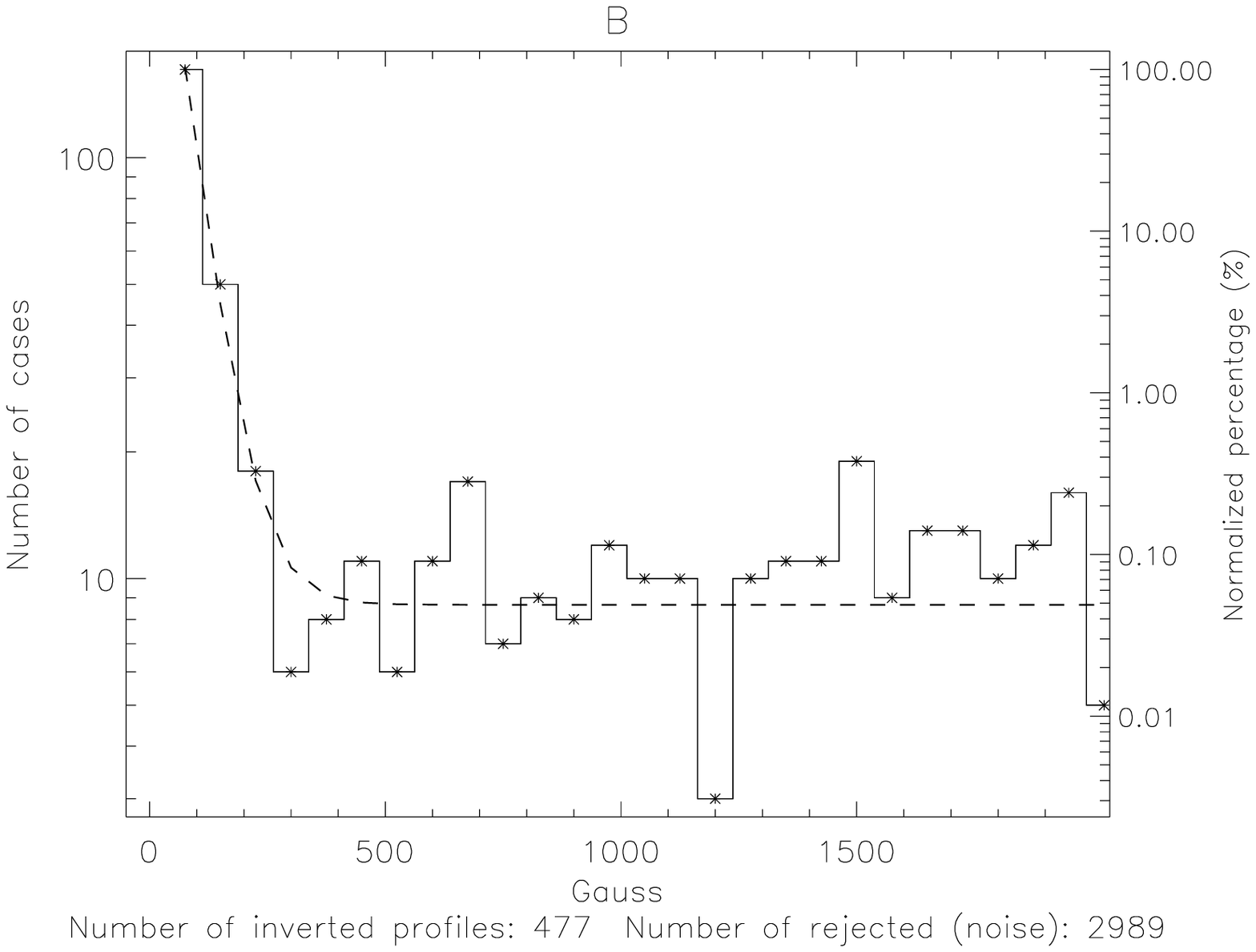}}
\resizebox{9cm}{!}{\includegraphics{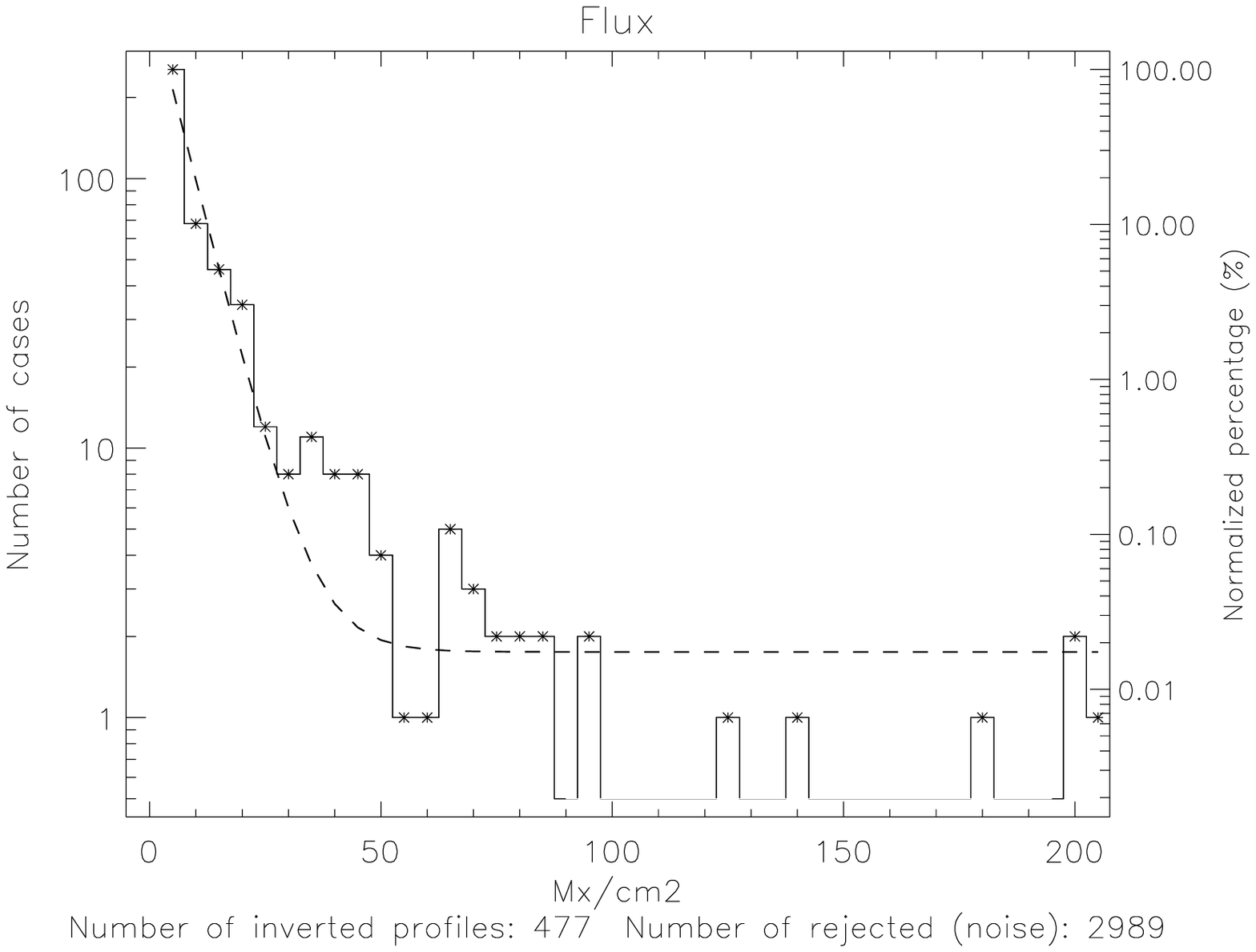}}
\resizebox{9cm}{!}{\includegraphics{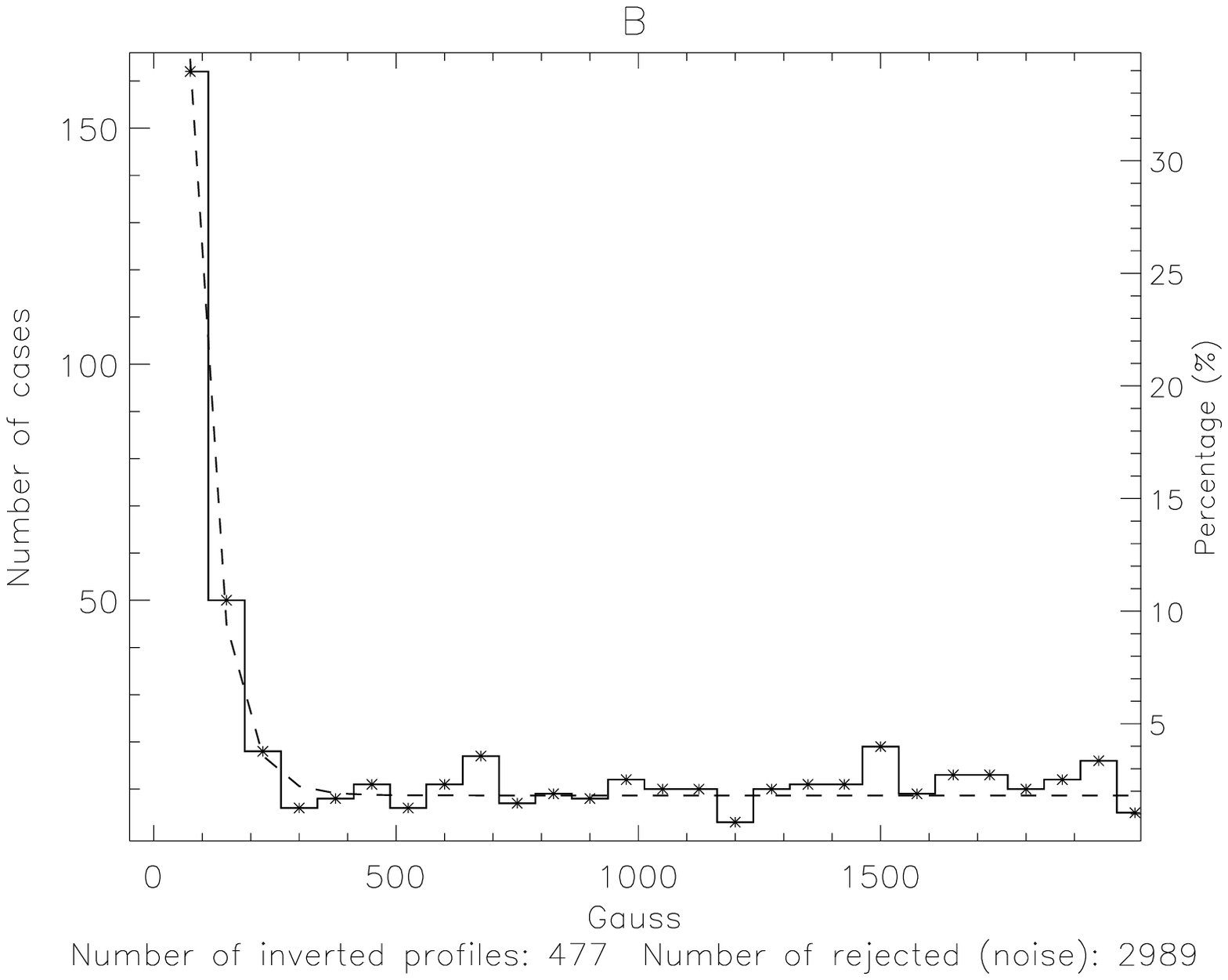}}
\resizebox{9cm}{!}{\includegraphics{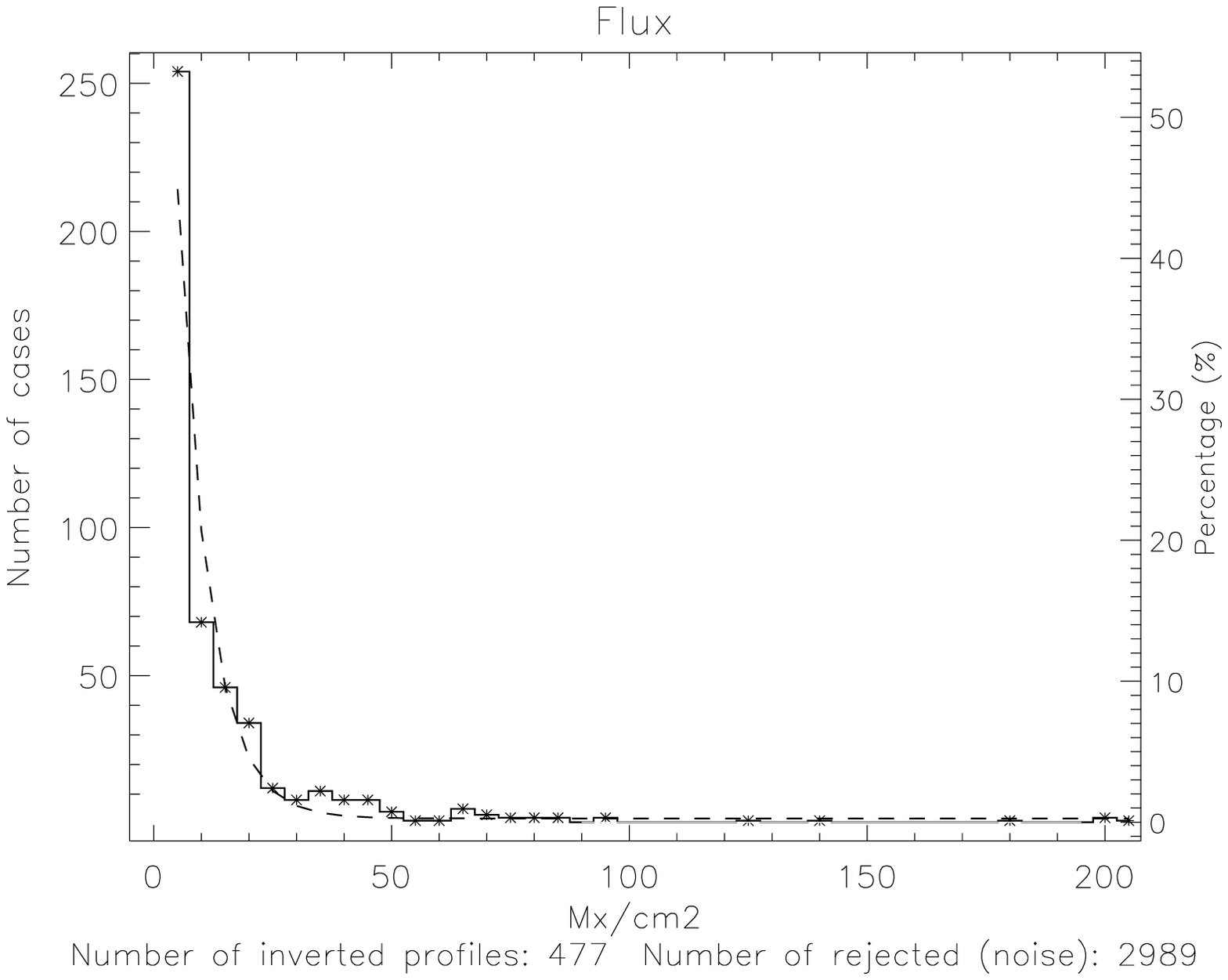}}
\caption{
\textit{Top panels:} 
Histograms of the distribution of the magnetic field strength and 
the longitudinal magnetic flux represented in logarithmic scale. 
\textit{Bottom panels:} The same histograms in normal scale.
The dashed lines represents the fitted   distribution 
functions N(x). The bin sizes are  75 G and 5 Mx/cm$^2$ in
field strength and  flux, respectively.  }
\label{fig:histos_b}
\end{figure*}

Only 14\%  of the total amount of data had a S/N ratio high enough to 
be inverted reliably: around 500 profiles.
In Fig. \ref{fig:histos_b}, we show the resulting  magnetic  
field strength and longitudinal  flux density distributions (defined as 
filling factor times the longitudinal field strength) from those inversions.
For better description, we have fitted the histograms to an empirical  distribution function of the form $ N(x)=a \exp(-x/b) + c$, where a, b and c are positives constants and N(x) represents the number of cases.   The introduction of a uniform distribution (the constant $c$) is required to fit the flat part of the histograms, corresponding to intermediate and high strength and flux density values. Otherwise, the left part of the histogram, associated to  weak strength and flux density values respectively,  is well fitted with an exponential function.
In the case of the field strengh distribution, the fitting  values are  a=672, b=51.3 G and c=8.6, while in the case of the flux density a=465, b=6.3  Mx/cm$^2$  and c=1.7. The full widths at half maximum are   respectively {\sc fwmh$_B$ = 54} G and {\sc fwmh}$_{Flux}$ = 4.6  Mx/cm$^2$, and the correspondent mean values are $<B>$ = 635 G and $<Flux>$ = 13 Mx/cm$^2$. 

Although one expects mostly maxwellian-like probability distribution functions (PDF) \cite{cerdenia06b} for the field strength, because of the coarseness of the histogram bins we prefer to fit the data with an exponential for the weak fields plus a constant distribution for the strong fields. In spite of that empirical fit, the histogram shown in Fig. \ref{fig:histos_b} should not be translated into that the most probable strength of the inverted profiles is zero G, but rather some value between $0<B<75$ G, included in the lower bin of the histogram. The main information the histograms show   is that the weak field strength regime (hG) is more abundant or dominates the surface coverage in the quiet sun conditions (70\%). Nevertheless, a considerable amount  of  cases,  30\%  of the inverted profiles, present magnetic strengths  superior to 1000 G, i.e. in the strong field strength regime, and cannot be considered as absent.

\subsection{Network versus Intranetwork}

\begin{figure*}[!htbp]
%%% two pannel graphs OVERPLOT
\resizebox{9cm}{!}{\includegraphics{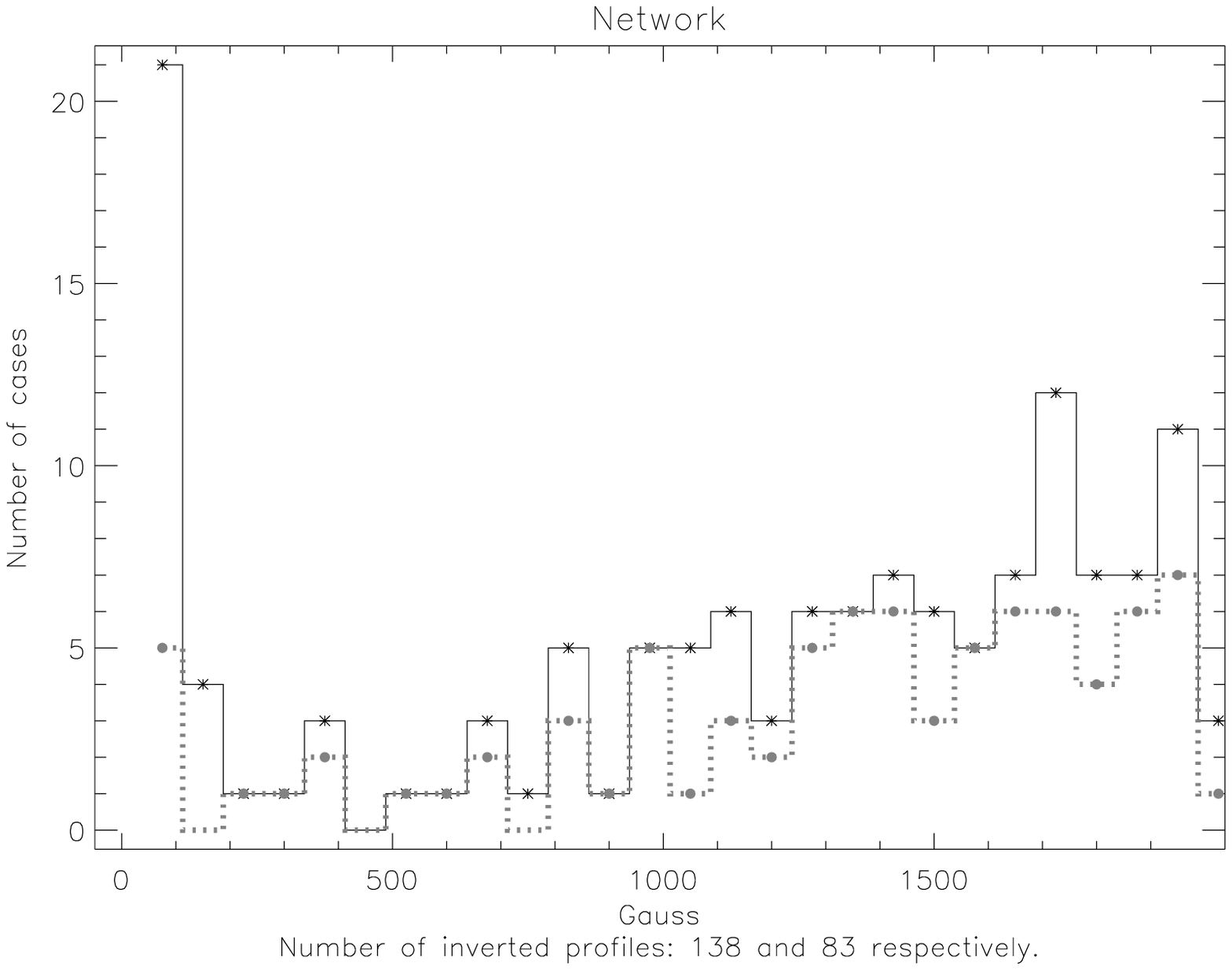}}
\resizebox{9cm}{!}{\includegraphics{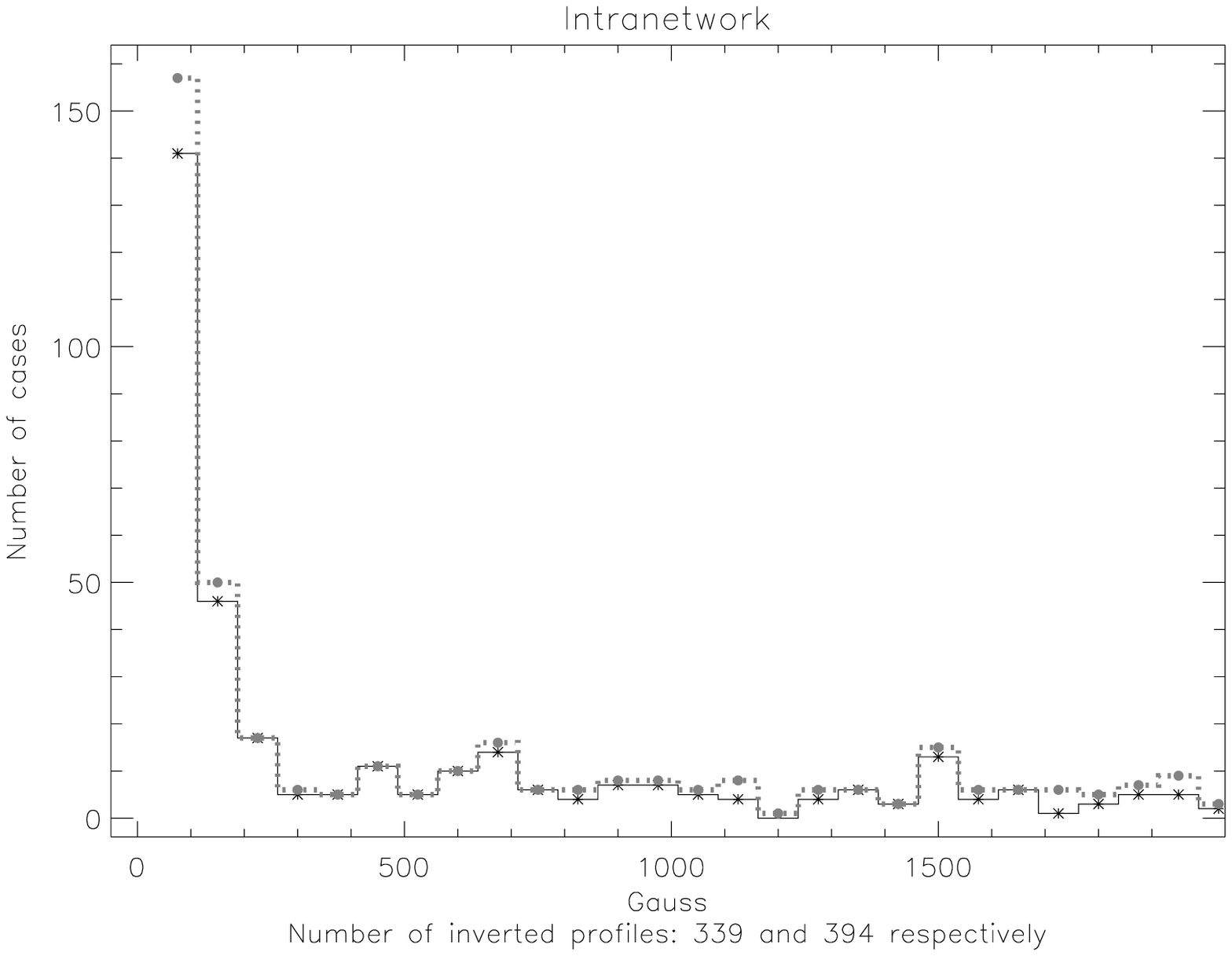}}
%%%%%%
\caption{Histograms of  network and intranetwork regions. The solid (dotted) line correspond to a threshold value of $1x10^{-3}$ ($2x10^{-3}$). In the network, the contribution of profiles with strength superior to 1 kG is considerably more important than in the intranetwork.}
\label{fig:histos_net}
\end{figure*}

In order to study how the presence of network and intranetwork regions influence the results shown in the histograms of Fig. \ref{fig:histos_b}, we proceed to differentiate the profiles between these two regions. To this purpose,  we have used the\- Fe {\sc i} line (5538.52 nm), sitting next to the Mn line, to set a criterion for the classification of profiles into one or the other regions. We arbitrarily set the following condition for points to belong to the photospheric network: If the level of the  normalized circular polarisation $V/I$ in the Fe {\sc i}  line exceeds a threshold value  ($TV$)  of the continuum level we consider the point as part of the network. All other points are  considered intranetwork.  We have initially used two threshold values:  $TV$= $1x10^{-3}$ and $2x10^{-3}$ . In  Fig. \ref{fig:histos_net}, we show both distributions (for field strength and longitudinal flux density) for each one of the regions. The solid line represent a  $TV$= $1x10^{-3}$ and the gray dotted line is for  $TV$= $2x10^{-3}$. 
When using the upper threshold, $TV$= $2x10^{-3}$ corresponding to higher flux concentrations, the contribution  of profiles in the weak strength regime in the network region has  almost disappeared while the number of profiles with strong field strengths has been just slightly  modified.
% This  indicates that if we continue increasing the threshold value,  the number of profiles belonging to the network (most %of them with kG strength fields) will decrease more and more without changing the main features. 
On the other hand, the use of the moderate threshold value   $1x10^{-3}$ already identifies the profiles in the strong regime in the network region as well as a non negligible contribution of fields with weak strengths.  In the intranetwork set, with either one or the other threshold  the general shape of the histogram  remains unchanged. It is clear that a different  threshold value able to radically change the general shapes of the histograms for both (network and intranetwork) regions would be unacceptably high and unrealistic. Therefore, and keeping in mind that small dependence of the network-assigned weak fields on the threshold value, we adopt
in what follows the network and intranetwork definitions using 
the lower TV, with correspondent distributions function represented by the solid line in Fig. \ref{fig:histos_net}. 
%Finally,  we know that any other higher TV used will just tend the analysis results to those obtained from the distribution in Fig. \ref{fig:histos_b}.}
To complete the justification on the criterion use for classification into network and internetwork,  let us add that from previous work \cite{LTC_hfs_06} we know that the described classification, however arbitrary, successfully identifies most of the network as identified visually over a magnetograph image. In terms of magnetic flux,  the threshold value of $1x10^{-3}$,   would correspond roughly to 10 Mx/cm$^2$, and it is certainly biased towards including any flux concentration found in the intranetwork into the network, and conversely, any low flux region in the network will be included in the intranetwork dataset, a bias that we should remember in the conclusions. 

After classification, from  a sample of 477 inverted cases, it was found that 70\%  belonged to the intranetwork region, while the resting 30\% were from the network.
%In figure \ref{fig:histos_net}, we show  the respective magnetic strength distributions for both regions. 
As already noticed in the previous paragraph, the correspondent distributions for the  magnetic field strength in the network region, represented by the solid line in the left panel in Fig. \ref{fig:histos_net}, clearly differs from the internetwork distribution and  also from the complete dataset distribution: %, shown in the bottom left panel of Fig. \ref{fig:histos_b}. 
 It presents an apparent growth in the number of cases in  the strong field regime, in a striking bimodal distribution.  On the other side, the intranetwork distribution keeps a similar shape to the one found before: The intranetwork appears to be predominantly permeated by weak fields.

The main conclusion arising from these separate distributions is that the derived magnetic strength  distribution from the inversions of the Mn {\sc i} line in the visible range  at 553 nm, in the low-flux quiet sun or intranetwork region (taking into account the known bias of our definition of intranetwork)  is dominated in number of cases by weak field-strength fields. This is in accord with  previous results obtained with the analysis by measurement  in the infrared range \cite{khomenko03, Lin95, LinRim99},  in the visible  \cite{keller_94} and with the recently retrieved distributions  from the very high spectral resolution (0.3 arc sec) {\sc hinode} data \cite{orozco_al_07}. Moreover, we found that the most probable expected strength is B $<$ 100 G, close to the  result obtained from simultaneous inversions from the infrared and from the visible recently published in  Mart\' inez Gonz\'alez et al. (2008).

On the other hand, strong flux regions, mostly the photospheric network, rather show a bimodal distribution of fields with a contribution of weak field-strength fields comparable to that found in weak-flux regions, and an increased appearance of kG fields \cite{cerdenia06b, lites_02}.

In the table \ref{tab:res_his} we summarise these results in terms of 
inversion percentages for different magnetic strengths as seen 
in the histograms of the analysed data sets previously discussed.

\begin{table}[ht]
\caption{Comparison between the total set of inverted profiles
and the Network-Intranetwork profiles.} 
\label{tab:res_his}
\centering
\begin{tabular}{|c|c|c|c|}
\hline 
&  Total inverted  & Intranetwork & Network \\
%\hline 
\hline 
Number of  Profiles & 477 & 339 & 138 \\
Strong regime  (kG) & 30\% & 17\% & 63\% \\
Weak regime (hG)    & 70\% & 83\% & 37\% \\
 $<$ B $>$ \ (G)    & 635  & 435  & 1130 \\
 $<$ Flux $>$ (Mx/cm$^2$) & 12.8  & 7.3  & 26.5 \\
\hline 
\end{tabular}
\end{table}

\section{Conclusions}
In this work  we contribute to  explore the diagnostic capabilities of Mn lines with strong hyperfine coupling for the magnetic topology of the quiet sun. We built an inversion code for the Mn line at 553.7 nm formed in a Milne-Eddington atmosphere, and tested it under the conditions typical of the quiet sun in terms of signal amplitudes and noise levels. Such conditions are highly constraining: in most of the cases the magnetic field information is contained exclusively in the Stokes V profile, the linear polarization signals being under noise levels and the intensity profile being dominated by the non-magnetic portion of the resolution element. Under such restrictive conditions the number of independent observables for the magnetic field is highly reduced \cite{ARetal07} and it sets a limit to the number of free parameters of our model. Previous works have shown that the Mn lines increased the number of observables thanks to the spectral features appearing under certain magnetic regimes. The inversion code we tested in this work takes advantage of it and allows us to improve over the simplistic weak-field approximation of previous works into a more sophisticated Milne-Eddington atmosphere and yet to conserve the possibility of disentangling strong from weak fields. 

Since most of the other parameters in the Milne-Eddington model, aside from field strength and flux, were not determined accurately under quiet-sun conditions, we doubt that any more sophisticated model can be used in the diagnostics at this point without the addition of further independent observables (e.g. with other spectral lines). However, more sophisticated models can shed light into implicit biases in such unrealistic picture of the solar photosphere as a Milne-Eddington atmosphere is, e.g. \cite{SA_hfs_08}.

We have payed special attention throughout this work to the numerical tests made to ascertain whether, under present observational conditions of the quiet sun, the weak fields were disentangled from the strong fields for similar net fluxes. The answer is affirmative and therefore we proceeded to apply the code to real data taken with the THEMIS telescope under good seeing conditions. Even in a telescope like THEMIS, focused on performing the most sensitive polarimetry possible, the acquisition of the data at the required signal-to-noise levels implied a trade-off in spatial resolution (seldomly better than 1 arc sec) that limits our results. We therefore constrained ourselves to separate the photospheric network from the intranetwork in our study, and restrained of identifying smaller scales, like granules and intergranules. 

To identify network from intranetwork we applied a rather poor rule fixing at $10^{-3}$ the smallest amplitude of the Stokes V profile respect to the continuum intensity for a point to belong to the photospheric network. Such a rule obviously classifies as network any high flux concentration, mostly found over the network, but not totally absent from the intranetwork. Conversely it calls intranetwork any low flux concentrations, mostly found all over the intranetwork but not absent from the network either. With such bias in mind we decided to keep the names of network and intranetwork for the two classes because a small change in such threshold value did not change the general shape of the histograms, but also because of the following reasons.

After inversion, the intranetwork points show a distribution of field strengths whose dominant feature can be easily fitted with an exponential tail. We use the noun \textit{tail} because an exponential distribution would result in a non-zero probability for null fields, something unacceptable for a vector field. Such a distribution, showing that the quiet sun regions are mostly permeated of weak strength fields, would be in full agreement with previous results in both Hanle and Zeeman effects \cite{manso04, LinRim99, khomenko03, MG_IRvsVis_08} and reveals a random or disorganised vector magnetic field at scales much smaller than present spatial resolution, if Hanle and Zeeman diagnostics are to be coherent with each other and with a field strength whose highest probability remains below 100G \cite{SAEC03}.

The network points, on the other hand, show a bimodal distribution: the exponential tail found in the intranetwork distribution is still present, peaking at field strengths below 100G. But on top of that disorganised magnetic regime we observe the appearance of stronger fields, in the kiloGauss regime. This second distribution of fields would be in agreement with the presence of one or several concentration mechanisms, able to stabilise, organise and perhaps amplify the statistical fluctuations of the ubiquitous turbulent field in those places where photospheric dynamics or sheer accidents are able to maintain a temporal coherence bigger than the typical magnetic diffussivities  \cite{parker_82}. One can heuristically think that the strong down drafting plumes in the vertex of the convection cells are such places \cite{nordlund92,rast_03}, thus defining the magnetic photospheric network as a non-continuous succession of bright network points \cite{MR_92} where magnetic fields are organised in vertical structures and made stable by the downdrafting plasma. Other smaller plumes found in mesogranular or granular scales \cite{rast_03} may also survive long enough to allow the statistical concentration of high fluxes, to be found here and there throughout the intranetwork and that, in our analysis, would have been included in the network group.

The cartoon just drawn above is in agreement with the histograms measured in the present work over the quiet sun. In particular, we are prone to  the presence of an ubiquitous random field, as inferred from Hanle diagnostics, and whose statistical fluctuations can be seen as Zeeman signatures over deep magnetograms, but otherwise highly disorganized at present resolutions except when plasma movements create conditions, as around convection plumes, stable enough to concentrate and build up net magnetic flux.  In any case, with the inversions performed here on the Mn {\sc i}  in the visible region of the spectra, at 553 nm, we have found that the distributions of the strength fields in the quiet Sun regions are dominated in occurrence by weak strength fields except in those areas related with the network.

\begin{acknowledgements}
  JCRV thanks to M.J. Mart\' inez Gonz\'alez for all helpful discussions and also to the referee, J. S\'anchez Almeida, for all the comments that served to improve the final state of the manuscript.  
\end{acknowledgements}

%%%%%%%%%%%%%%%%%%%%%%% BIBLIO %%%%%%%%%%%%%%%%%%%%%%%
\bibliographystyle{/home/julio/articulos/bibtex/aa}

\end{document}